\documentclass[onecolumn,12pt,showpacs,amsmath,amssymb]{revtex4}

\usepackage{graphicx}
\usepackage{dcolumn}
\usepackage{bm}

\begin{document}

\title{Coulomb crystals in the magnetic field}

\author{D.~A.\ Baiko}
\affiliation{A.F.\ Ioffe Physical-Technical Institute,
     Politekhnicheskaya 26, 194021 St.-Petersburg, Russian Federation}

\date{\today}

\begin{abstract}
The body-centered cubic Coulomb crystal of ions in the presence of 
a uniform magnetic field is studied using the rigid electron
background approximation. The phonon mode spectra are calculated for a wide
range of magnetic field strengths and for several orientations of
the field in the crystal. The phonon spectra are used to calculate
the phonon contribution to the crystal energy, entropy, 
specific heat, Debye-Waller
factor of ions, and the rms 
ion displacements from
the lattice nodes for a broad range of densities, temperatures, chemical
compositions, and magnetic fields. Strong magnetic field
dramatically alters the properties of quantum crystals. The phonon
specific heat increases by many orders of magnitude. The ion displacements
from their equilibrium positions become strongly anisotropic. 
The results can be relevant for dusty plasmas, ion plasmas in Penning
traps, and especially for the crust of magnetars (neutron stars with
superstrong magnetic fields $B \gtrsim 10^{14}$~G). The effect of
the magnetic field on ion displacements in a strongly magnetized
neutron star crust can suppress the nuclear reaction rates and make them
extremely sensitive to the magnetic field direction.
\end{abstract}


\maketitle

\section{Introduction}
\label{introduct}
The model of a crystal of point-like charges immersed in 
a uniform neutralizing background of opposite charge was conceived by Wigner
\cite{W34} to describe a possible crystallization of electrons.
These Wigner crystals of electrons have much in common with Coulomb 
crystals of ions with the uniform electron background. 
The model of the Coulomb crystal is widely used in
different branches of physics, including theory of plasma oscillations 
(e.g., \cite{P65}), solid state physics,
and works on dusty plasmas and ion plasmas in Penning traps
(e.g., \cite{IBTJHW98}).

Moreover, Coulomb crystals of ions, immersed in an almost uniform
electron background, are formed in the cores of white dwarfs and
envelopes of neutron stars. The properties of such crystals are
important for the structure and evolution of these astrophysical 
objects (e.g., \cite{CADW92,HPY07}). 
In particular, the Coulomb crystal heat capacity  \cite{BPY01} 
controls cooling of old white dwarfs
and is used to determine their ages (e.g., \cite{WKCM09}).
Crystallization of white dwarfs can influence their pulsation
frequencies. It can thus be studied by powerful methods of
asteroseismology \cite{MMK04}. Microscopic properties of
Coulomb crystals determine the
efficiency of electron-phonon scattering in white dwarfs and neutron stars, 
and, hence, transport
properties of their matter (such as electron thermal and electric
conductivities, and shear viscosity, e.g., \cite{BKPY98,CY05}) as well as the
neutrino emission in the electron-ion bremsstrahlung process
\cite{KPPTY99}. Many-body ion correlations in dense matter produce 
screening of ion-ion (nucleus-nucleus) Coulomb interaction and affect
nuclear reaction rates in the thermonuclear burning regime with
strong plasma screening and in the pycnonuclear burning regime (when
the reacting nuclei penetrate through the Coulomb barrier owing to
zero-point vibrations in crystalline lattice). The description of
various nuclear burning regimes and observational manifestations of
burning in white dwarfs and neutron stars  are discussed in
Refs.\ \cite{SVH69,YGABW06,CD09} and references therein. The manifestations
include type Ia supernova explosions of massive accreting white
dwarfs, bursts and superbursts, deep crustal heating of accreted
matter in neutron stars.

Since the late 1990s certain astrophysical applications have been requiring
a comprehensive study of Coulomb crystals in strong magnetic fields.
The topic has become important due to
the growing observational evidence that some very intriguing
astrophysical objects, soft-gamma repeaters (SGRs) and anomalous
X-ray pulsars, belong to the same class of sources called magnetars
(see, e.g., \cite{T02,K04,WT06} and reference therein).
These are thought to be isolated, sufficiently warm neutron stars
with extremely strong magnetic fields $B \gtrsim 10^{14}$~G. For
instance, the magnetic field of SGR 1806--20, inferred from
measurements of its spin-down rate, is $B \sim 2 \times 10^{15}$ G
\cite{catalog}. Magnetars are observed in all ranges of the
electromagnetic spectrum.
They show powerful quasipersistent X-ray
emission, bursts and giant bursts with enormous energy release.
During giant bursts, one often observes quasi-periodic X-ray
oscillations which are interpreted (e.g., \cite{WS07}) as vibrations
of neutron stars (involving torsion vibrations of crystalline
neutron star crust). It is likely that the activity of
magnetars is powered by their superstrong magnetic fields. 
Thus the magnetars can be viewed as natural laboratories of 
dense matter in magnetic fields. 
In order to build adequate 
models of magnetar envelopes and
interpret numerous observations, it is crucial to know the properties of
magnetized Coulomb crystals.
The main goal of this paper is to study in detail
Coulomb crystals in an external uniform magnetic field.
(The results reported here were partially 
presented in \cite{B00}.)

The Coulomb crystals in question consist of fully ionized ions with
charge $Ze$ and mass $M$ arranged in a crystal lattice and immersed 
into the rigid
electron background 
(in this case, rigid means unpolarizable or incompressible, 
i.e.\ constant and uniform). 
The effect of the magnetic field $B$ on the ion
motion can be characterized by the ratio
\begin{equation}
   b = \omega_B / \omega_{p},
\label{rhoB}
\end{equation}
where
\begin{equation}
  \omega_B=\frac{ZeB}{Mc}, \quad
  \omega_{p}=\sqrt{\frac{4 \pi Z^2 e^2 n}{M}}
\end{equation}
are the ion cyclotron frequency and the ion plasma frequency,
respectively; $n$ is the ion number density, while $c$ is the speed
of light. It is expected that the magnetic field modifies the
properties of the ion crystal at $b \gtrsim 1$ (see, however, Figs.\
\ref{b-c} and \ref{u1um1} below). In a strong magnetic field the
approximation of rigid electron background is a bigger idealization
of the real situation in neutron star crust matter, since higher densities are
required to achieve full ionization and suppress the polarizability
of the electron background. The higher densities (and $\omega_{p}$)
imply smaller $b$. However, the effective ion charge approximation
turns out to be successful for analyzing partially ionized systems
(e.g., \cite{PCY97}). The quantity $b$ (also equal to
$v_{A}/c$, where $v_{A} = B/\sqrt{4 \pi \rho}$ is the Alfv\'en
velocity, $\rho$ being the mass density) is, actually, 
independent of the ion charge. Hence, one can consider large $b$ in
Coulomb crystals at not too high densities having in mind the
effective ion charge approximation. Despite that, the effect of 
the polarizability of the compensating
electron background has to be studied separately.

The closely related problem of 
magnetized Wigner crystals of electrons 
was studied in the early 1980s by Usov,
Grebenschikov and Ulinich \cite{UGU80} and by Nagai and Fukuyama
\cite{NF82, NF83}. Usov et al.\ \cite{UGU80} obtained the equations
for crystal oscillation modes, studied qualitatively the oscillation
spectrum, and diagonalized the Hamiltonian of the crystal for a
proper quantum description of the oscillations in terms of phonons.
In addition, the authors investigated asymptotic temperature and
magnetic field dependences of the specific heat, rms electron
displacement from the respetive lattice site, and
magnetic moment of the crystal. They obtained a soft phonon mode
with the frequency $\Omega \propto k^2$ near the center of the
Brillouin zone. It resulted in a rather unusual low-temperature
specific heat behavior ($T^{3/2}$) instead of the standard Debye law
($T^3$). Also, the authors mentioned the dependence of the crystal
energy on the magnetic field direction as well as the increased
stability of the crystal due to a suppression of electron vibration
amplitudes. In addition, Usov et al.\ \cite{UGU80} considered the
dielectric function of the crystal and studied the effect of the 
electromagnetic
field induced by the electron motion. Their main results were, however,
of semi-qualitative nature, limited to various extreme cases, whereas
the present paper focuses mostly on quantitative results, pertaining to
Coulomb crystals of ions, with an eye to astrophysical
implications.

Nagai and Fukuyama \cite{NF82} calculated phonon spectra of
magnetized body-centered cubic and face-centered cubic (bcc and fcc)
Wigner crystals and compared the energies of these crystals at zero
temperature as a function of the magnetic field. The crystal energy
was calculated as a sum of the electrostatic (Madelung) energy,
and the zero-point energy of crystal vibrations. The effect of
the anharmonic and exchange terms was neglected. The electrostatic
energy is independent of the magnetic field (as long as the field
does not alter the lattice structure). The vibration energy does depend
on the field because the field modifies phonon modes. At zero field 
the energy minimum is realized by the bcc
structure, both with and without zero-point vibrations (in general,
it is not true for polarizable background, e.g., \cite{B02}).
The authors showed that for a sufficiently strong magnetic field and
at relatively high densities [$r_s  < 275$, where $r_s = a_e/a_0$ is
the standard density parameter, $a_e=(3/4 \pi n_e)^{1/3}$, and $a_0$
is the Bohr radius, $n_e$ being the electron number density] the
full energy is minimized by the fcc structure. It is worth
mentioning that the authors did not consider the dependence of the
zero-point energy on the magnetic field direction and performed all
calculations for a fixed direction (along one of the high symmetry
axes of the crystals). However, this choice of the field direction
for the bcc lattice was not optimum, as far as the lattice
energy was concerned. Moreover, the energy gain obtained by choosing
the optimum direction is of the same order of magnitude as the
difference between the zero-point energies of bcc and fcc lattices.

Nagai and Fukuyama \cite{NF83} investigated an analogous structural
transition between bcc and hexagonal close-packed (hcp) electron
Wigner lattices. The hcp lattice was found energetically favorable
for a sufficiently strong magnetic field at $r_s < 10700$. This
result seems more robust since the energy difference between bcc and
hcp lattices is several times larger than the energy gain obtained
by choosing the optimum direction of the magnetic field in the bcc
lattice (the field direction adopted in \cite{NF83} was the same as
in \cite{NF82}).

In addition, Nagai and Fukuyama \cite{NF83} described the behavior
of all 6 phonon modes of the hcp lattice in the magnetic field. Also, they
analyzed quantitatively the behavior of transverse and longitudinal
electron displacements from the equilibrium positions for bcc and hcp
lattices (at zero temperature) and concluded (qualitatively) that in
strong magnetic fields the crystals became significantly more stable
in the transverse direction and somewhat more stable in the
longitudinal direction.

The present paper is organized as follows. Section \ref{eqs}
discusses equations for Coulomb crystal oscillations 
in the magnetic field.
Section \ref{spectrum} focuses on the phonon spectrum of
such a crystal with a simple lattice (i.e., one ion in the primitive
cell). The Hamiltonian of the system is diagonalized in Sec.\
\ref{diag}, which allows one to express the ion displacement
operator via phonon creation and annihilation operators. Section
\ref{thermodyn} presents numerical calculations of thermodynamic
functions of the Coulomb bcc crystal for a wide range of densities,
temperatures and magnetic fields.
In Sec.\ \ref{crust_ex} these results are applied to the real
physical system found in magnetar crust.
Section \ref{rms} is devoted to an analysis of the 
rms amplitudes of ion vibrations and
to the Debye-Waller factor of the magnetized Coulomb crystal. 
Finally, Sec.\
\ref{moments} considers the dependence of phonon spectrum moments on
the magnitude and direction of the magnetic field.

The results will be parameterized by the quantum parameter
$\theta=\hbar \omega_{p}/T \equiv T_{p}/T$ (where $T$ and $T_{p}$
are temperature and ion plasma temperature, respectively), and by
the Coulomb coupling parameter $\Gamma = Z^2 e^2 /(aT)$. In this
case, $a=(3/4 \pi n)^{1/3}$ is the ion sphere radius. The Boltzmann
constant is set equal to $k_{B}=1$.

\section{Dispersion equation in the magnetic field}
\label{eqs}
%
Let us consider a Coulomb crystal
 of ions in a uniform magnetic field
$\bm{B}$. The Lagrangian of an ion is
\begin{equation}
      L =  \frac{M v^2}{2} + \frac{Ze}{c} \,
      \bm{A} \cdot \bm{v} - Ze \,\phi~,
\label{1p-lagr}
\end{equation}
where $\bm{v}$ is the ion velocity, $\bm{ A}$ is the vector
potential, and $Ze \,\phi$ is the potential energy of the ion
($\phi$ being the scalar potential). Choosing the vector potential
acting on the $i$-th ion as $A_i^\alpha = [\bm{ B} \times \bm{
u}_i]^\alpha/2 = \varepsilon^{\alpha \beta \gamma} B^\beta
u_i^\gamma/2$ (where $\bm{ u}_i$ is the $i$-th ion displacement), 
the Lagrangian of the ion system can be written as:
\begin{eqnarray}
          L_B &=& L_{0} + \frac{Z e}{c} \sum_{i=1}^N A_i^\alpha
   \dot{u}_i^\alpha
\nonumber \\
 &=& L_{0} + \frac{M \omega_B}{2} \sum_{i=1}^N
       \varepsilon^{\alpha \beta \gamma} \dot{u}_i^\alpha n^\beta
u_i^\gamma~.
\label{LB-ini}
\end{eqnarray}
In this case, $\bm{n}$ is the unit vector along the magnetic field,
$N$ is the total number of ions,
and $L_0$ is the field-free Lagrangian.

Introducing the same collective coordinates as at $\bm{B}=0$,
\begin{eqnarray}
        u_{{\rm l}p}^\alpha &=& \sqrt{\frac{N_{\rm cell}}{M N}}
        \sum_{\bm{ k}}
      {\cal A}^\alpha_{\bm{ k}p} \exp{\left(i\bm{k} \cdot \bm{ R}_{\rm l} 
      \right)}~,
\label{uofA-2} \\
        {\cal A}^\alpha_{\bm{ k}p} &=& \sqrt{\frac{M N_{\rm cell}}{N}}
        \sum_{{\rm l}} u_{{\rm l}p}^\alpha
        \exp{\left( -i \bm {k}\cdot \bm{ R}_{\rm l}\right)}~,
\label{Aofu-2}
\end{eqnarray}
one can rearrange the Lagrangian as
\begin{equation}
        L = L_0 + \frac{\omega_B}{2}
        \sum_{\bm {k}, p}
     \varepsilon^{\alpha \beta \gamma}
      \dot{{\cal A}}^\alpha_{ \bm{ k}p}  n^\beta
    {\cal A}_{-\bm{ k}p}^\gamma~.
\label{LofAgen-2}
\end{equation}
In this case, $\bm{k}$ is a wavevector in the first Brillouin zone,
index {\rm l} enumerates direct lattice vectors $\bm{R}_{\rm l}$,
$N_{\rm cell}$ is the number of ions in the primitive cell ($N_{\rm
cell}=1$ for simple lattices), and index $p$ goes from 1 to $N_{\rm
cell}$. Summations in Eqs.\ (\ref{uofA-2}) and (\ref{Aofu-2}) are
over all wavevectors in the first Brillouin zone and over all
direct lattice vectors, respectively. In the thermodynamic limit,
with ${\cal A}^\alpha_{\bm{ k}p}$ being interpreted as generalized
coordinates of the system, one obtains the following Euler equation:
\begin{equation}
       \ddot{{\cal A}}^\alpha_{\bm{ k}p} +
        \sum_{p'}
       {\cal D}_{pp'}^{\alpha \beta}(\bm{ k})
        {\cal A}^\beta_{\bm{ k}p'} +
 \omega_B \varepsilon^{\alpha \beta \gamma} n^\beta
\dot{{\cal A}}^\gamma_{\bm{ k}p}= 0~,
\label{Eulereq-2}
\end{equation}
where ${\cal D}_{pp'}^{\alpha \beta}(\bm{ k})$ is the dynamic
matrix of the lattice at $\bm{B}=0$.
The Fourier-transform
${\cal A}^\alpha_{\bm{ k}p}(t) \to A^\alpha_{\bm{k}p}(\Omega)$
yields the algebraic equation
\begin{equation}
       - \Omega^2 A^\alpha_{\bm{k}p} +
        \sum_{p'}
       {\cal D}_{pp'}^{\alpha \beta}(\bm{k})
        A^\beta_{\bm{k}p'} -i \Omega \omega_B
\varepsilon^{\alpha \beta \gamma} n^\beta A^\gamma_{\bm{k}p} = 0~,
\label{omegEuler-2}
\end{equation}
that can be solved if the ion vibration frequency $\Omega$ satisfies
the dispersion equation below \cite{UGU80}
\begin{equation}
       {\rm det} \left\{ {\cal D}_{pp'}^{\alpha \beta} -
\Omega^2 \delta_{pp'} \delta^{\alpha \beta} - i \Omega \omega_B
     \varepsilon^{\alpha \gamma \beta} n^\gamma  \right\} =0~.
\label{secular-2}
\end{equation}
Due to the presence of the third term on the left-hand side, Eq.\
(\ref{omegEuler-2}) does not represent the eigennumber problem for a
Hermitian matrix, and the respective polarization vectors are not
orthogonal. This complicates the reduction of the Hamiltonian to the
sum of Hamiltonians of independent oscillators. The proper procedure
will be discussed in Sec.\ \ref{diag}.

\section{Phonon spectrum}
\label{spectrum}
There are $3 N_{\rm cell}$ oscillation modes
for a given vector $\bm{k}$
in the first Brillouin zone. The frequencies of these modes satisfy 
the generalized
Kohn's sum rule 
$\sum_s \Omega^2_{\bm{k}s} = N_{\rm cell} (\omega_{p}^2  + \omega_B^2)$
\cite{NF82}.

At small $k = |\bm{k}|$ the behavior of phonon frequencies is more
complex than without the field. It depends substantially on the
direction of $\bm{k}$ with respect to $\bm{B}$. We restrict
ourselves to $N_{\rm cell} = 1$. It is possible to obtain the exact
asymptotes of $\Omega$ at small $k$ (cf.\ Ref.\ \cite{NF82}). In
this limit, the dispersion equation (\ref{secular-2}) can be written
as:
\begin{equation}
   - \frac{\Omega^6}{\omega^6_p} + (1+ b^2) \,\frac{\Omega^4}{\omega^4_p}
   + E_B \, \frac{\Omega^2}{\omega^2_p} + F_0 = 0~,
\label{sec-ass-2}
\end{equation}
where
\begin{eqnarray}
       16 \pi E_B &=& 16 \pi E_0 -
       16 \pi b^2 (\hat{\bm{k}} \cdot \bm{n})^2
\nonumber \\
       &+& (ka_{l})^2 b^2 [(\beta + \gamma_2) +
(\alpha+2 \gamma_2) (\hat{\bm{k}} \cdot \bm{n})^2
\nonumber \\
 &+& (\gamma_1 -3 \gamma_2)
(n_x^2 \hat{k}_x^2 + n_y^2 \hat{k}_y^2 +n_z^2 \hat{k}_z^2)]~,
\label{BB}
\end{eqnarray}
$\hat{\bm{k}}=\bm{k}/k$,
while the coefficients $E_0$ and $F_0$ are the same as
in the field-free dispersion equation \cite{CK57}:
\begin{eqnarray}
       8 \pi E_0 &=& (ka_{l})^2 [(\beta + \gamma_2) 
                       +  (\gamma_1 - 3 \gamma_2)
                           (\hat{k}_x^2 \hat{k}_y^2 + \hat{k}_y^2 \hat{k}_z^2 
                    + \hat{k}_z^2 \hat{k}_x^2)]~,
\nonumber \\
       256 \pi^2 F_0 &=& (ka_{l})^4 [(\beta + \gamma_2)^2 
                       + 2 (\beta + \gamma_2) (\gamma_1 - 3 \gamma_2)
                           (\hat{k}_x^2 \hat{k}_y^2 + \hat{k}_y^2 \hat{k}_z^2 
                    + \hat{k}_z^2 \hat{k}_x^2) 
\nonumber \\
                       &+& 3 (\gamma_1 - 3 \gamma_2)^2 \hat{k}_x^2 
                    \hat{k}_y^2 \hat{k}_z^2]~.
\label{C0}
\end{eqnarray}
The constants $\alpha$, $\beta$, and $\gamma_{1,2}$ characterize the
crystal at $\bm{B}=0$. They were defined and calculated in
Ref.~\cite{CK57}, and were recalculated for bcc and fcc Coulomb
lattices in Ref.~\cite{B00}. They are reproduced in Table
\ref{constants}, along with the lattice constant $a_{l}$, for
completeness. The subscripts $x$, $y$, and $z$ refer to the
Cartesian coordinate system aligned with the main cube of the
respective reciprocal lattice. 

The asymptote of the smallest
frequency $\Omega_1$ at $k \to 0$ can be derived by dropping the
$\Omega^6$-term and choosing the smallest root of the remaining
quadratic equation:
\begin{equation}
        \frac{\Omega^2_1}{\omega_{p}^2} = \frac{1}{2 (1+ b^2)}
  \left[-E_B-\sqrt{E^2_B-4(1+ b^2) F_0} \right]~.
\label{ass-o1}
\end{equation}
At sufficiently small $k$, $\Omega_1^2/\omega_{p}^2 \approx
F_0/|E_B|$. Then, for the phonons, propagating strictly
perpendicular to the magnetic field, $\Omega_1 \propto k$,
and the phonons are acoustic.
If, on the other hand, $\bm{k}\cdot \bm{ n} \ne
0$, then $\Omega_1 \propto k^2$ at small $k$, in contrast with the
linear dependence at $\bm{B}=0$. As the angle between $\bm{k}$ and
$\bm{B}$ decreases, the quadratic asymptote of $\Omega_1$ becomes
valid for wider range of $k$. 
Neglecting angular dependences and numerical factors in 
Eqs.\ (\ref{BB}) and (\ref{C0}),
one can estimate
the lowest phonon frequency as  
$\Omega_1/\omega_p \sim  k a_l / \sqrt{2+b^2}$ for
propagation perpendicular to the field,
and as 
$\Omega_1/\omega_p \sim k^2 a_l^2 / b$
for $\bm{k}\cdot \bm{n} \ne 0$.
In both cases, at $b \gg 1$, the mode typical energy
is inversely proportional to the field strength. 

\begin{table}[ht]
\caption[]{Dynamic matrix coefficients for bcc and fcc lattices}
\begin{eqnarray}
\begin{tabular}{c|ccccc}
    & $\alpha$ & $\beta$ & $\gamma_1$ & $\gamma_2$ & $n a_{l}^3$  \\ \hline
  ~~~bcc~~~ & ~$4.1243864$~ & ~$0.84911538$~ & ~$-3.4886939$~ & 
                                                 ~$-1.5915193$~ & ~$2$~ \\
  ~~~fcc~~~ & ~$4.0036483$~ & ~$1.2376805$~ &  ~$-4.2930041$~ & 
                                                 ~$-1.71184285$~ & ~$4$~ \\
\end{tabular}
\nonumber
\end{eqnarray}
\label{constants}
\end{table}

The asymptote of the biggest frequency $\Omega_3$ can be found by
dropping the last term in Eq.\ (\ref{sec-ass-2}) and choosing the
maximum root of the remaining quadratic equation:
\begin{equation}
        \frac{\Omega^2_3}{\omega^2_p} =
    \frac{1}{2}\left[1+b^2 +  \sqrt{(1+b^2)^2 +
    4 E_B}\right]~.
\label{ass-o3}
\end{equation}
At small $k$ this yields
\begin{eqnarray}
    \frac{\Omega_3^2}{\omega_{p}^2} &=&
    \frac{1}{2} \left[  1+b^2 
  +  \sqrt{(1+b^2)^2 - 4 b^2
  (\hat{\bm{k}}\cdot \bm{ n})^2}
                 \right] + O(k^2)~.
\label{ass-o3-k0}
\end{eqnarray}
In general, at any $\bm{k}$ and at $b
\gg 1$, the biggest frequency $\Omega_3 \approx \omega_B$. This
corresponds to the conventional cyclotron ion motion.

Consider $k=0$. Then $\Omega_3^2/\omega_{p}^2 = 1+b^2$ for phonons
propagating perpendicular to the field. From the sum rule, it follows that
the intermediate frequency $\Omega_2$ becomes 0 at $k=0$. Because
$\Omega_2 > \Omega_1$, it is clear that $\Omega_2 \propto k$ for $k \to
0$. If, on the other hand, the phonon propagates along the magnetic
field, then at $k=0$ one has $\Omega_3^2/\omega_{p}^2 =
\max{(1,b^2)}$, and, hence, $\Omega_2^2/\omega_{p}^2 =
\min{(1,b^2)}$. Therefore both maximum and intermediate frequency
modes are optical. As the angle between $\bm{k}$ and $\bm{B}$
decreases from $\pi/2$ to 0, the value of $\Omega_2$ at $k \to 0$ 
increases from 0 to $\min{(\omega_p,\omega_B)}$, whereas the 
value of $\Omega_3$ at 
$k \to 0$  decreases from $\sqrt{\omega_p^2+\omega_B^2}$ to 
$\max{(\omega_p,\omega_B)}$.

Obviously, Eq.\ (\ref{sec-ass-2}) can be solved analytically without
neglecting any terms. However, at very small $k$ analytical schemes
become numerically unstable. The thermodynamic properties of Coulomb
crystals at low temperatures require integration of functions
containing the phonon frequencies near the center of the Brillouin
zone. Under these conditions and especially at high magnetic fields
the asymptotes given here become helpful.

\begin{figure}[t]
\begin{center}
\leavevmode
\includegraphics[height=73mm,bb=71 532 540 740,clip]{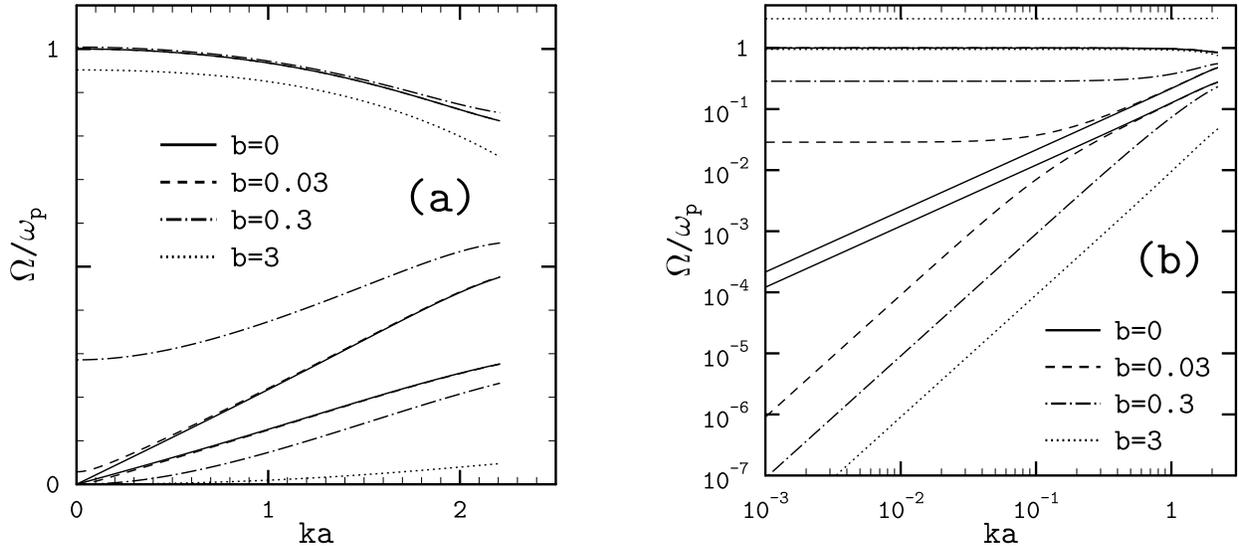}
\end{center}
\vspace{-0.4cm} \caption[ ]{Phonon spectrum of the magnetized bcc
lattice for several values of $b = \omega_B/\omega_{p}$ in natural
(a) and logarithmic (b) scales. Slowly growing
mode $\Omega/\omega_{p} \approx 3$ at $b=3$ is not
shown in panel (a). Magnetic field and wavevector $\bm{k}$ orientations 
are defined in the text; $a$ is the ion sphere radius.} 
\label{spectinb}
\end{figure}
%
%

Figure \ref{spectinb} shows the phonon spectrum of the bcc lattice
(in natural and logarithmic scales) in the direction of $\bm{k}$
determined by the polar angle $\theta = 1.3 \tan^{-1}{\sqrt{2}}$ and
the azimuthal angle $\phi=0.9 \pi/4$ [angles $\theta$ and $\phi$ are
defined with respect to the same Cartesian reference frame as the
$x,y,z$-subscripts in Eqs.\ (\ref{BB}) and (\ref{C0})]. The magnetic
field is parallel to the direction from an ion towards one of its
nearest neighbors ($\theta =\tan^{-1}{\sqrt{2}}$, $\phi=\pi/4$). For
$ka \lesssim 0.01$, analytic asymptotes discussed above are used; at
higher $ka$, exact calculations are performed. One can observe the
quadratic dependence of the lowest frequency on $k$ near the
Brillouin zone center. The dependence becomes linear closer to the
Brillouin zone boundary (at $b \ll 1$).

The polarization of the electron background in the absence of the
magnetic field converts the optical mode with a frequency $\approx
\omega_{p}$ into an acoustic mode in the vicinity of the Brillouin
zone center (e.g., \cite{PH73,B02}). The background polarization
effects in a magnetized crystal will be investigated elsewhere. A
similar conversion effect can be expected in this case as well, except that
one of the modes must remain optical close to the center of the
zone, with a frequency $\Omega \approx \omega_B$. This is so, because
the coefficient of the $\Omega^4$-term in the dispersion equation
(\ref{secular-2}), that determines the sum of the squares of all
frequencies, will continue to contain $\omega_B^2$, whereas in the
field-free case at $k \to 0$ the $\Omega^4$-term tends to zero for
polarizable background.

\section{Hamiltonian of the Coulomb crystal in the magnetic field}
\label{diag}
In this section the Hamiltonian of the Coulomb crystal in the
magnetic field is represented as a sum of Hamiltonians of
independent oscillators. The derivation follows that of Ref.\
\cite{UGU80}, but some additional details are provided. Consider the
Cartesian reference frame with the axes directed along the
eigenvectors of the matrix ${\cal D}^{\alpha \beta}(\bm{ k})$ (again
$N_{\rm cell}=1$). Then the Lagrangian (\ref{LofAgen-2}) has the
form
\begin{eqnarray}
        L_B &=& \frac{1}{2} \sum_{\bm{k}}
          \dot{{\cal A}}^\alpha_{\bm{k}}\,
         \dot{{\cal A}}^\alpha_{-\bm{k}} - U_0 -
         \frac{1}{2}
         \sum_{\bm{k}} \omega^2_{\bm{k} \alpha}\,
         {\cal A}^\alpha_{\bm{k}}\,
         {\cal A}^\alpha_{-\bm{k}}
\nonumber \\
  &+&
         \frac{\omega_B}{2}
        \sum_{\bm{k}}
     \varepsilon^{\alpha \beta \gamma}
      \dot{{\cal A}}^\alpha_{\bm{k}} \, n^\beta \,
    {\cal A}_{-\bm{k}}^\gamma~.
\end{eqnarray}
In this case, $\omega_{\bm{k} \alpha}$ is the phonon frequency at
$\bm{B}=0$. Omitting the constant equilibrium electrostatic energy
term $U_0$ and turning to the Hamiltonian $H_B$, one has
\begin{eqnarray}
       {\cal P}^\alpha_{\bm{k}} &=& \dot{{\cal A}}^\alpha_{-\bm{k}} +
        \frac{\omega_B}{2} \varepsilon^{\alpha \beta \gamma} n^\beta\,
     {\cal A}_{-\bm{k}}^\gamma~,
\label{calP} \\
       H_{B} &=& \sum_{\bm{k}} \dot{\cal A}^\alpha_{\bm{k}}\,
     {\cal P}^\alpha_{\bm{k}} - L_B
\nonumber \\
         &=& \frac{1}{2} \sum_{\bm{k}}
   \pi^\alpha_{-\bm{k}}\, \pi^\alpha_{\bm{k}} +
    \frac{1}{2} \sum_{\bm{k}} \omega^2_{\bm{k} \alpha}\,
        {\cal A}^\alpha_{\bm{k}}\,
         {\cal A}^\alpha_{-\bm{k}}~,
\label{HB}
\end{eqnarray}
where $\pi^\alpha_{\bm{k}}={\cal P}^\alpha_{\bm{k}} - \omega_B
\varepsilon^{\alpha \beta \gamma} n^{\beta} {\cal
A}_{-\bm{k}}^{\gamma}/2$. In the quantum formalism, ${\cal
A}^\alpha_{\bm{k}}$ and ${\cal P}^\alpha_{\bm{k}}$ become operators
with the usual commutation rules $\left[{\cal
A}^\alpha_{\bm{k}},{\cal P}^\beta_{\bm{ k'}} \right] = i \hbar
\delta^{\alpha \beta} \delta_{\bm{k}{\bm{ k'}}}$. However, the
operators $\pi^\alpha_{\bm{k}}$ and $\pi^\beta_{-\bm{k}}$ do not
commute at $\alpha \ne \beta$: $\left[\pi^\alpha_{\bm{k}},
\pi^\beta_{{\bm{ k}'}} \right] = i \hbar \omega_B \varepsilon^{\beta
\gamma \alpha} n^\gamma \delta_{{\bm{ k}'}, -\bm{k}}$. Consequently,
the creation and annihilation operators, defined in the same way as
at $\bm{B}=0$, but with $\pi^\alpha_{\bm{k}}$ in place of ${\cal
P}^\alpha_{\bm{k}}$, do not satisfy the required commutation
relationships.

In this situation one constructs the creation and annihilation
operators as linear combinations of the operators of generalized
coordinates and momenta (e.g., Ref.\ \cite{UGU80}),
\begin{equation}
        \hat{a}^\dagger_{\bm{k}} = \alpha^\lambda_{\bm{k}} 
        \pi^\lambda_{\bm{k}}
   + \beta^\lambda_{\bm{k}}  {\cal A}^\lambda_{-\bm{k}}~,
\label{anz-a+}
\end{equation}
where $\alpha^\lambda_{\bm{k}}$ and $\beta^\lambda_{\bm{k}}$ are
constant coefficients (and summation over $\lambda$ is implied).
They can be determined from the equation $\left[H_B,
\hat{a}^\dagger_{\bm{k}} \right] = \hbar \Omega_{\bm{k}}
\hat{a}^\dagger_{\bm{k}}$ (e.g., Ref.\ \cite{UGU80}). Using
Eqs.~(\ref{HB}) and (\ref{anz-a+}), one arrives at
\begin{eqnarray}
          \hbar \Omega_{\bm{k}} \,
      (\alpha^\lambda_{\bm{k}}\, \pi^\lambda_{\bm{k}}
   + \beta^\lambda_{\bm{k}} \, {\cal A}^\lambda_{-\bm{k}}) &=&
      i \hbar \omega_B \, \varepsilon^{\lambda \gamma \alpha} n^\gamma
        \alpha^\lambda_{\bm{k}}\, \pi^\alpha_{\bm{k}}
\nonumber \\
              &+&
     i\hbar \omega^2_{\bm{k}\lambda} \alpha^\lambda_{\bm{k}}\,
        {\cal A}^\lambda_{-\bm{k}}
       - i \hbar \beta^\lambda_{\bm{k}} \,
\pi^\lambda_{\bm{k}}~,
\label{su-gen}
\end{eqnarray}
or
\begin{eqnarray}
        \Omega_{\bm{k}} \, \beta^\lambda_{\bm{k}} &=& i
         \omega^2_{\bm{k}\lambda} \, \alpha^\lambda_{\bm{k}}~,
\label{su-b} \\
        \Omega_{\bm{k}} \, \alpha^\lambda_{\bm{k}} &=&
       i  \omega_B \, \varepsilon^{\alpha \gamma \lambda} n^\gamma
        \alpha^\alpha_{\bm{k}} - i  \beta^\lambda_{\bm{k}}~,
\label{su-a}
\end{eqnarray}
so that
\begin{equation}
        (\omega^2_{\bm{k}\lambda} - \Omega_{\bm{k}}^2)
          \, \alpha^\lambda_{\bm{k}} -
        i \Omega_{\bm{k}} \, \omega_B \,
        \varepsilon^{\lambda \gamma \alpha} n^\gamma
        \, \alpha^\alpha_{\bm{k}} = 0~.
\label{Eq-for-al}
\end{equation}
The system (\ref{Eq-for-al}) can be solved only if the quantities
$\Omega_{\bm{k}}$ satisfy the dispersion equation (\ref{secular-2}),
i.e., if they are the eigenfrequencies of the oscillation modes with the
wavevector $\bm{k}$. Therefore, there are three possible solutions
for the operator $\hat{a}^\dagger_{\bm{k}}$, corresponding to the
three generally different frequencies $\Omega_{\bm{k}s}$, $s=1,2,3$:
$\hat{a}^\dagger_{\bm{k}s} = \alpha^\lambda_{\bm{k}s}
(\pi^\lambda_{\bm{k}} + i \omega^2_{\bm{k}\lambda} {\cal
A}^\lambda_{-\bm{k}} / \Omega_{\bm{k}s})$. Evidently, the equations
for the coefficients $\alpha^\lambda_{\bm{k}s}$ and
$\alpha^\lambda_{-\bm{k}s}$ coincide. Thus from now on index $\bm{k}$
will be omitted where possible.

The solutions of the system (\ref{Eq-for-al}) are sets of the
cofactors to any of the rows of the matrix \cite{UGU80}
\begin{eqnarray}
 \left\Vert
\begin{array}{rrr}
   \omega^2_{x} - \Omega_{s}^2 &
    i \Omega_{s} \omega_B n^z &
   - i \Omega_{s} \omega_B n^y \\
   - i \Omega_{s} \omega_B n^z &
   \omega^2_{y} - \Omega_{s}^2 &
    i \Omega_{s} \omega_B n^x \\
   i \Omega_{s} \omega_B n^y &
   - i \Omega_{s} \omega_B n^x &
   \omega^2_{z} - \Omega_{s}^2 \\
\end{array} \right\Vert~
\end{eqnarray}
(these sets are proportional to each other).
Now one has to normalize these solutions together with the operators
$a^\dagger_{s}$, so that
$\left[\hat{a}_{s}, \hat{a}^\dagger_{s}  \right] =1$. Accordingly,
\begin{eqnarray}
          \left[\hat{a}_{s}, \hat{a}^\dagger_{s'} \right] &=&
       \alpha_s^{\lambda \ast} \alpha_{s'}^{\lambda'} i \hbar \omega_B
       \varepsilon^{\lambda' \gamma \lambda} n^\gamma
\nonumber \\
       &+& \alpha_s^{\lambda \ast} \alpha_{s'}^{\lambda}
             \hbar \omega^2_{\lambda}
      \left( \frac{1}{\Omega_{s}} +
      \frac{1}{\Omega_{s'}} \right)
\nonumber \\
   &=& \alpha_s^{\lambda \ast} \alpha_{s'}^{\lambda} \hbar
    \left( \frac{\omega^2_{\lambda}}{\Omega_{s}} +
         \Omega_{s'} \right)~,
\label{ort-gen}
\end{eqnarray}
where Eq.\ (\ref{Eq-for-al}) was employed. Now consider the
following sequence of equations,
\begin{eqnarray}
     0 &=& \alpha_{s'}^{\lambda} \,[( \Omega^2_{s} -
               \omega^2_{\lambda} )\,
       \alpha_s^{\lambda \ast} -
         i \Omega_{s}
      \omega_B \varepsilon^{\lambda \gamma \nu} n^\gamma \alpha_s^{\nu \ast} ]
\nonumber \\
    &=&  \alpha_{s'}^{\lambda}\, \alpha_s^{\lambda \ast} ( \Omega^2_{s}
            - \omega^2_{\lambda} )
          + \alpha_{s'}^{\nu}\, \alpha_s^{\nu \ast}
           \frac{\Omega_{s}}{\Omega_{s'}} \,
           (\omega^2_{\nu} - \Omega^2_{s'})
\nonumber \\
      &=& \alpha_{s'}^{\lambda}\, \alpha_s^{\lambda \ast}
    \left(\frac{\Omega_{s}}{\Omega_{s'}} -1 \right)
     (\omega^2_{\lambda} + \Omega_{s}
      \Omega_{s'})
\label{aux-2}
\end{eqnarray}
[where Eq.\ (\ref{Eq-for-al}) was used twice]. Thus the
right-hand side of Eq.~(\ref{ort-gen}) vanishes, if $\Omega_{s} \ne
\Omega_{s'}$. This property is analogous to the orthogonality
condition for the polarization vectors at $\bm{B}=0$. At $s=s'$, 
Eq.\ (\ref{ort-gen}) determines the normalization of the coefficients
$\alpha^\lambda_s$ :
\begin{equation}
        \alpha^\lambda_s \, \alpha^{\lambda \ast}_s \hbar
     \left(\frac{\omega^2_{\lambda}}{\Omega_{s}} +
         \Omega_{s} \right) = 1~.
\label{norma}
\end{equation}
All the other commutators of the operators
$\hat{a}_{s}$ and $\hat{a}^\dagger_{s}$
vanish automatically.

Owing to the equality $\left[H_B, \hat{a}^\dagger \right] =
\hbar \Omega \hat{a}^\dagger$ and the normalization
relationships obtained above, the Hamiltonian (\ref{HB})
can be rewritten in the canonical form
\begin{equation}
       H_B = \sum_{\bm{k}s} \hbar \Omega_{\bm{k}s}
       \left( \hat{a}^\dagger_{\bm{k}s} \hat{a}_{\bm{k}s}
          + \frac{1}{2}\right).
\label{HBofa}
\end{equation}
Moreover, with the help of Eqs.\ (\ref{HB}) and (\ref{anz-a+}), it 
is possible to prove 
the relationships:
\begin{eqnarray}
        \sum_s \alpha^\lambda_s \, \alpha^{\lambda' \ast}_s
        - \alpha^{\lambda \,\ast}_s \, \alpha^{\lambda'}_s &=& 0,
\label{o1} \\
      \hbar  \sum_s \frac{\omega_\lambda \omega_{\lambda'}}{\Omega_s}
      (\alpha^\lambda_s \, \alpha^{\lambda' \ast}_s
        + \alpha^{\lambda \,\ast}_s \,\alpha^{\lambda'}_s) &=&
          \delta^{\lambda \lambda'},
\label{o2} \\
       \hbar \sum_s \Omega_s
      (\alpha^\lambda_s \,\alpha^{\lambda' \ast}_s
        + \alpha^{\lambda\, \ast}_s \,\alpha^{\lambda'}_s) &=&
          \delta^{\lambda \lambda'}~.
\label{o3}
\end{eqnarray}
Multiplying $\hat{a}^\dagger_{-\bm{k}s}$ by $\alpha_{\bm{k}s}^{\lambda \ast}$,
and $\hat{a}_{\bm{k}s}$ by $\alpha_{\bm{k}s}^{\lambda}$, summing over $s$, 
subtracting
the second expression from the first one, and also using (\ref{o1}) and
(\ref{o2}), one obtains
\begin{equation}
          {\cal A}^\lambda_{\bm{k}} =-i \hbar
      \sum_s \, \alpha_{\bm{k}s}^{\lambda \,\ast}\,
          \hat{a}^\dagger_{-\bm{k}s} +
         i \alpha_{\bm{k}s}^{\lambda} \,\hat{a}_{\bm{k}s}~.
\label{Aofa-2}
\end{equation}
The above formula is instrumental in calculations of such crystal properties
as the Debye-Waller factor and the rms
ion displacement from a lattice site in the magnetic field.
These results are reported in Sec. \ref{rms}.

\section{Phonon thermodynamic functions of the magnetized Coulomb crystal}
\label{thermodyn}
Phonon thermodynamic functions in the magnetic field are calculated using
the same general formulas (e.g., \cite{LL84}) and numerical
integration schemes (e.g., \cite{AG81, BPY01, B00}) as in the
field-free case. The phonon free energy (with phonon chemical potential 
$\mu=0$ and neglecting zero-point contribution) reads
\begin{eqnarray}
      F &=& T 
   \sum_{\bm{k}s} \ln{\left[1-\exp{\left(-\frac{\hbar \Omega_{\bm{k}s}}{T}
        \right)}\right]} 
\nonumber \\
        &=&  V \sum_s \int_{\rm BZ} \frac{{\rm d} \bm{k}}{(2 \pi)^3}
        \ln{\left[1-\exp{\left(-\frac{\hbar \Omega_{\bm{k}s}}{T}\right)}
        \right]}~,  
\label{Fph}
\end{eqnarray}
where $V$ is the volume, and the integral is over the first Brillouin zone.
The phonon thermal energy $E$ and heat capacity $C$  
are then given by
\begin{eqnarray}
          E &=& F - T \left(\frac{\partial F}{\partial T}\right)_{\mu, V} 
          = \sum_{\bm{k}s} \frac{\hbar \Omega_{\bm{k}s}}{e^{\hbar 
            \Omega_{\bm{k}s}/T}-1}
\label{E/N} \\
          C &=& - T \left(\frac{\partial^2 F}{\partial T^2}\right)_{\mu, V} 
             = \frac{1}{4T^2} \sum_{\bm{k}s} 
          \frac{\hbar^2 \Omega_{\bm{k}s}^2}{{\rm sinh}^2(\hbar 
            \Omega_{\bm{k}s}/2T)}~.
\end{eqnarray}
In the magnetic field, there are two new
parameters, $b = \omega_B/\omega_{p}$ and the field direction. A
study of frequency moments of the phonon spectrum in Sec.\
\ref{moments} will show that the dependence of these moments on the
field direction is rather weak. This allows one to expect that the
dependence of thermodynamic functions on the field direction is
also weak. Thus, in the present section the consideration is
restricted to the field direction, which corresponds to the minimum
zero-point energy of the crystal. For the bcc lattice, it is the
direction from a lattice site to one of its closest neighbors.

\begin{figure}[t]
\begin{center}
\leavevmode
\includegraphics[height=73mm,bb=71 532 540 740,clip]{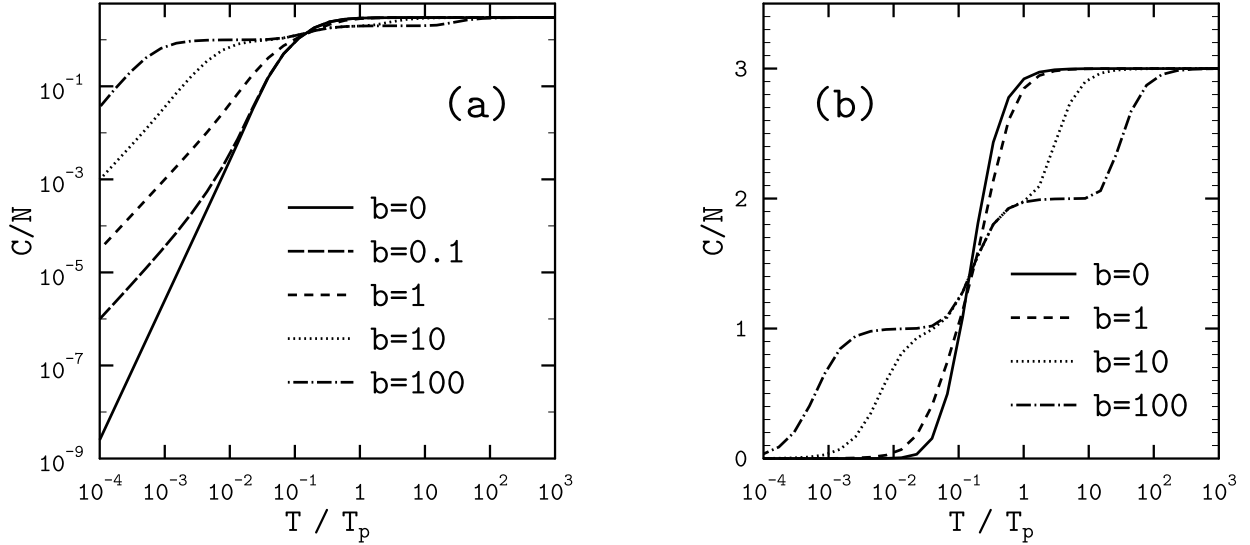}
\end{center}
\vspace{-0.4cm} \caption[ ]{Phonon specific heat (per ion) for the
bcc lattice as a function of temperature for several values of the
magnetic field in logarithmic (a) and linear (b) 
scales. The magnetic field is directed towards one of the
nearest neighbors. } \label{b-c}
\end{figure}
%
%

The calculated phonon heat capacity per one ion is presented in
Fig.\ \ref{b-c} in logarithmic and linear scales [panels (a) and (b),
respectively] as a function of $\theta^{-1}=T/T_{p}$ for
several values of $b$. In Fig.\ \ref{b-c}(a) 
one can clearly see the
change of the low-temperature asymptote from $T^3$ to $T^{3/2}$ due
to the appearance of the soft mode $\Omega \propto k^2$ for non-zero
$B$. Also, this easily excited mode (e.g., at $b=100$) is responsible
for a relatively high specific heat $C/N \sim 1$ all the way up to
$\theta \sim 10^3$. At these temperatures the field-free specific
heat is already down by 6 orders of magnitude.

In strong magnetic fields, higher temperatures are required to
achieve the classical regime as compared to the field-free case.
This is so, because the classical regime occurs when many phonons are
excited in {\it all} modes. In a strong magnetic field, there is the
high-frequency (cyclotron) mode, $\Omega_3 \approx \omega_B$ (Sec.\
\ref{spectrum}). Hence, the classical regime is realized if $b
\theta \ll 1$, in contrast to the conventional criterion $\theta \ll
1$ at $B=0$. This is illustrated in Fig.\ \ref{b-c}(b). 
For instance, for $b = 100$ the classical value $C/N = 3$
is reached only at $\theta \lesssim 0.01$.

\begin{figure}[ht]
\begin{center}
\leavevmode
\includegraphics[height=73mm,bb=33 27 344 344,clip]{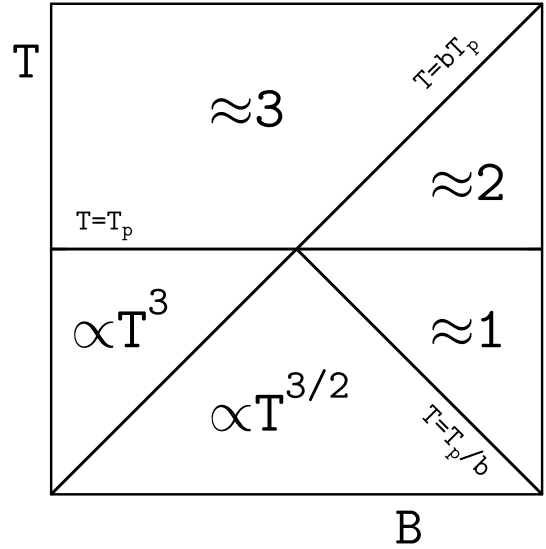}
\end{center}
\vspace{-0.4cm} \caption[ ]{A sketch of the specific heat behavior 
in various domains 
of the $T$-$B$ plane. The point, where all straight lines cross, 
corresponds to $b=\theta=1$.} 
\label{tb-diag}
\end{figure}
%
%

The energies of three phonon modes are spaced far away from each
other in strong magnetic fields. The minimum frequency $\Omega_1
\propto 1/B$, the maximum frequency $\Omega_3 \approx \omega_B$, and
the intermediate frequency $\Omega_2 \sim \omega_{p}$ 
(Sec.\ \ref{spectrum}). This gives
rise to the pronounced staircase structure of the heat capacity seen
in Fig.\ \ref{b-c}(b) at $b=10$ and $100$.

This discussion is further illustrated by Fig.\ \ref{tb-diag}, where
one can assess qualitatively the behavior of the specific heat in various 
domains of the $T$-$B$ plane. At high temperatures, the crystal is classic
and the specific heat reaches its maximum value of 3. 
In strong magnetic fields ($b \gg 1$) when the temperature drops below 
$\hbar \omega_B$, the cyclotron phonon mode is frozen out and 
$C/N \approx 2$. As the temperature drops further, below $T_p$, the
intermediate mode $\Omega_2$ freezes out in large portions of the
Brillouin zone, and the specific heat, now mainly due to the 
fully-excited soft mode, approaches 1. 
By contrast, in a non-magnetized ($b \ll 1$) crystal, the two lower
modes deviate from being acoustic only in the very vicinity of the
Brillouin zone center, while the third one is $\approx \omega_p$. 
Hence, when $T$ drops below $T_p$ in such a crystal,
the Debye law $C \propto T^3$ is recovered. Finally, at any $b$, 
when the temperature is so low, that only the least energetic 
mode contains any heat and the $\Omega \propto k^2$ law is probed, 
$C \propto T^{3/2}$ behavior results.

\begin{figure}[ht]
\begin{center}
\leavevmode
\includegraphics[height=73mm,bb=71 532 540 740,clip]{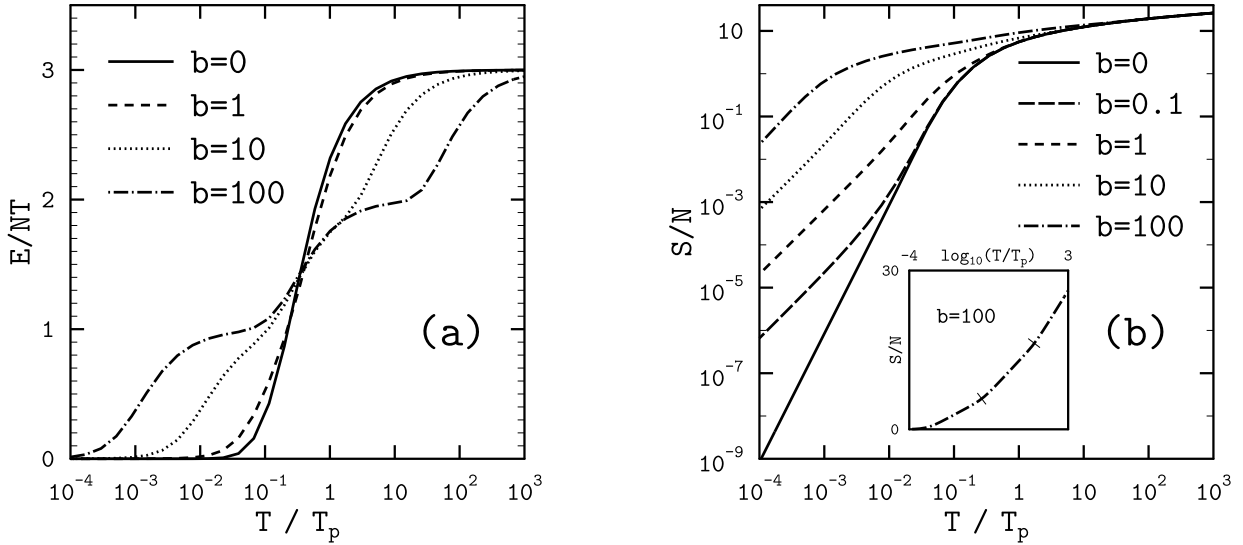}
\end{center}
\vspace{-0.4cm} \caption[ ]{Phonon energy per ion divided by $T$
(a) and phonon entropy per ion (b) for the
bcc lattice as functions of temperature for several values of the
magnetic field. Inset in panel (b) shows $S/N$ in the 
linear scale at $b=100$.
The magnetic field is directed towards one of the
nearest neighbors.} 
\label{b-se}
\end{figure}
%
%

In Fig.\ \ref{b-se} phonon thermal energy and entropy $S=(E-F)/T$ 
are plotted.
In the case of energy, plotted in the linear scale in Fig.\ \ref{b-se}(a), 
one observes 
again the staircase structure. Though less pronounced, 
it occurs for the same reason
as for the specific heat. The classical harmonic oscillator limit
$E=3NT$ is naturally reproduced. At low temperatures, the magnetic field 
gives rise to the $T^2$ dependence of the energy instead of the
$T^4$ field-free asymptote. For sufficiently
low temperatures, the thermal energy is dominated by the zero-point energy, 
determined by the spectrum moment $u_1$ and discussed 
in Sec.\ \ref{moments}. The numerical calculations of the entropy
are shown in Fig.\ \ref{b-se}(b). In this case,
as for the specific heat, the familiar field-free $T^3$ asymptote
is replaced by $T^{3/2}$ in the quantum magnetized crystal. 
In the classical regime, the entropy 
depends on temperature logarithmically 
and is insensitive to the magnetic field. 
This asymptote is discussed in some detail in Sec.\ \ref{moments}.
When drawn in linear scale, the entropy, as a function of $\log_{10}{T}$, 
also shows
an atypical structure at $b \gg 1$ [inset in Fig.\ \ref{b-se}(b)]. 
In this case, the dependence becomes
piece-wise linear with several slope changes,
corresponding to the sequential excitation of the three phonon modes.

Clearly, magnetized crystal thermodynamics, constructed in this Section,
cannot be described by the well-known in 
solid state physics Debye model (e.g., \cite{LL84}). On the other hand, 
at $b \gg 1$, partial thermodynamic quantities due to the cyclotron 
mode $\Omega_3$, can be represented
with high accuracy by the simple Einstein model (e.g., \cite{BH54}).

\section{Application to magnetar crust}
\label{crust_ex}
\begin{figure}[ht]
\begin{center}
\leavevmode
\includegraphics[height=73mm,bb=13 9 350 344,clip]{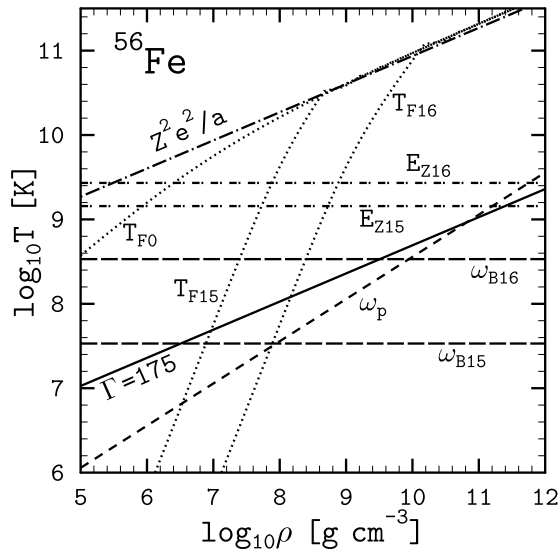}
\end{center}
\vspace{-0.4cm} \caption[ ]{Temperature-density diagram for fully
ionized $^{56}$Fe matter. $\Gamma=175$ is the classical crystal
melting line. $T_{\rm F}$ and $E_{\rm Z}$ mark electron degeneracy
temperature and electron-ion binding energy (see text for details).
Subscripts 0, 15, and 16 refer to the magnetic field values $B=0$,
$10^{15}$, and $10^{16}$ G, respectively. Note, that 
$E_{{\rm Z}0} = 1.1 \times 10^8$ K.} \label{fe-phys}
\end{figure}

As a practical application of the previous section results, 
consider fully ionized iron plasma in neutron star crust 
(Sec.\ \ref{introduct}). 
Typical values of $\theta$ and $b$ 
at $B=10^{15}$ and $10^{16}$~G can be estimated from Fig.\ \ref{fe-phys}.   
The curve $\Gamma=175$ shows the melting line of
the classical Coulomb crystal at ${B}=0$. The line $T=Z^2 e^2 /a$
represents typical electrostatic energy per ion (and separates the
regions of a free ion gas above the line, and a strongly coupled ion
system below the line). The electron degeneracy temperature is shown
by dotted lines marked $T_{\rm F0}$, $T_{\rm F15}$, and $T_{\rm
F16}$ for three values of the magnetic field, $B=0$, $10^{15}$, and
$10^{16}$ G, respectively. Finally, $E_{\rm Z15}$ and $E_{\rm Z16}$
are the electron ground-state energies of one-electron iron ion for
$B=10^{15}$ or $10^{16}$ G, respectively. These energies are
calculated using rescaling of equivalent hydrogen energies $E (Z,B)
= Z^2 E (Z=1,B/Z^2)$ (e.g., \cite{WRH81}), and the fitting
formula \cite{P98} for the energy of the $s=0$ state in the hydrogen
atom. The assumption of full ionization in the degenerate plasma can
be used if $T_{{\rm F}} > E_{\rm Z}$ (for a given magnetic field).
At $T_{\rm F} \gg E_{\rm Z}$, the electron gas is nearly
incompressible, and the
rigid electron background model is very well
justified. At $T_{{\rm F}} \lesssim E_{\rm Z}$ the plasma cannot be
treated as fully ionized, but even in this case one can use
present results for estimates by employing the effective ion charge
approximation.

%
%

\begin{figure}[ht]
\begin{center}
\leavevmode
\includegraphics[height=73mm,bb=71 532 540 740,clip]{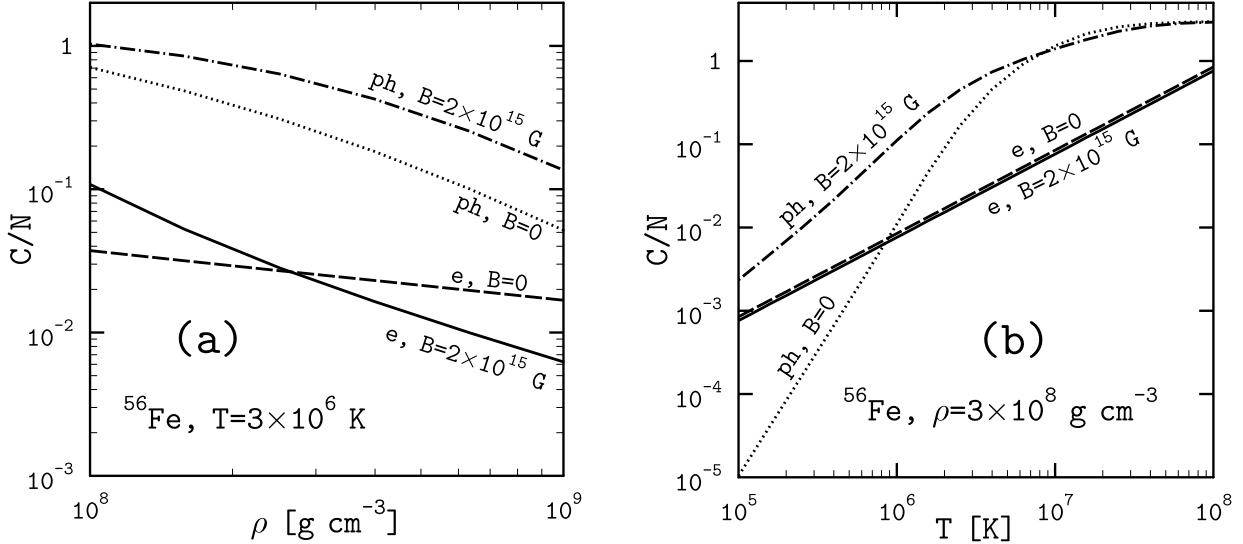}
\end{center}
\vspace{-0.4cm} \caption[ ]{Phonon and electron specific heat (per
ion) of fully ionized $^{56}$Fe matter versus density (a)
and temperature (b) at $B=0$ and $2 \times 10^{15}$ G.
For the $B \neq 0$ case, $b \theta = 23$ in panel (a); 
$T_p = 6.2 \times 10^7$ K and $b=1.09$ in panel (b).}
\label{iron_spec_heat}
\end{figure}

Figure \ref{iron_spec_heat} demonstrates the phonon and electron
specific heat (per ion) of fully ionized $^{56}$Fe plasma as a function
of density (a) and temperature (b) at $B=0$ and $2
\times 10^{15}$ G. The electron contribution is calculated using
standard formulas for strongly degenerate relativistic Fermi gas.
For $B=2 \times 10^{15}$~G under the displayed conditions the plasma
electrons fill the lowest Landau level only; the next Landau level
would be occupied at $\rho \gtrsim 10^9$ g cm$^{-3}$, and the plasma
would become partly ionized at $\rho \lesssim 10^8$ g cm$^{-3}$
(cf.\ Fig.\ \ref{fe-phys}). The temperature dependence of the
electron specific heat is linear, and near coincidence of two
electron lines in Fig.\ \ref{iron_spec_heat}(b) is
largely accidental. For the ions to crystallize, the temperature
must be below the melting temperature, which is $\sim 10^8$ K
in the density range considered (cf.\ Fig.\ \ref{fe-phys}).

%
%

The phonon contribution at $B=0$, recalculated here, reproduces the
results of \cite{BPY01}. At $B=2\times 10^{15}$ G the results
of the present work are plotted.
As seen from Fig.\ \ref{iron_spec_heat}, phonons dominate electrons
in the specific heat in a wide range of temperatures and densities.
The magnetic field provides an extra boost to the phonon specific
heat, especially at lower temperatures (in the quantum regime) due
to the easily excited soft mode.

\section{Ion vibrations and the Debye-Waller factor
in the magnetized crystal}
\label{rms}
Using Eqs.\ (\ref{uofA-2}) and (\ref{Aofa-2}), one can derive the
expression for the operator of the ion displacement from 
its equilibrium lattice position:
\begin{eqnarray}
   \hat{u}_{\rm l}^\lambda = \frac{i \hbar}{\sqrt{MN}}\;
        \sum_{\bm{k}s} (\alpha_{\bm{k}s}^\lambda \,\hat{a}_{\bm{k}s} -
        \alpha_{\bm{k}s}^{\lambda\, \ast} \,\hat{a}^\dagger_{-\bm{k}s})
        \exp{\left(i\bm{k} \cdot \bm {R}_{\rm l}\right)}~.
\label{uofa-2}
\end{eqnarray}
Then the rms ion displacement
$r_T = \sqrt{\langle \hat{u}^\lambda \, \hat{u}^\lambda \rangle_T}$
can be calculated as
\begin{eqnarray}
         \frac{r^2_T}{a^2} &=& \frac{\hbar^2}{MN a^2} \sum_{\bm{k}s}
            \alpha_{\bm{k}s}^{\lambda \, \ast}\, 
             \alpha_{\bm{k}s}^{\lambda}\,
         \langle \hat{a}^\dagger_{\bm{k}s} \, \hat{a}_{\bm{k}s} +
        \hat{a}_{\bm{k}s}\, \hat{a}^\dagger_{\bm{k}s} \rangle_T
\nonumber \\
    &=& \frac{\theta}{3 N \Gamma} \sum_{\bm{k}s}
         ( \hbar \omega_{p} \alpha_{\bm{k}s}^{\lambda \, \ast} \, 
             \alpha_{\bm{k}s}^{\lambda} ) \,
            (2 \bar{n}_{\bm{k}s} + 1) \equiv \frac{\theta}{\Gamma} \, 
            {\cal F}\left(\theta,b\right)
\label{r2T}
\end{eqnarray}
[cf.\ Eq.\ (\ref{norma})],
where $\bar{n}_{\bm{k}s} = (e^{\hbar \Omega_{\bm{k}s}/T} - 1)^{-1}$
is the mean number of phonons in a mode ${\bm{k}s}$. 
In addition, it is possible to find the rms ion
displacement $u_{\hat{\bm {q}}}$ in the direction along an arbitrary
unit vector $\hat{\bm {q}}$,
\begin{eqnarray}
     u^2_{\hat{\bm {q}}} &=&
    \langle \hat{u}^\lambda \, \hat{u}^\mu \, \hat{q}^\lambda \,
             \hat{q}^\mu \rangle_T
\nonumber \\
&=&  \frac{\hbar^2}{MN} \sum_{\bm{k}s}
            \alpha_{\bm{k}s}^{\lambda \, \ast}\, \alpha_{\bm{k}s}^{\mu} \,
        \hat{q}^\lambda \, \hat{q}^\mu \,
            (2 \bar{n}_{\bm{k}s} + 1)~.
\label{u2q}
\end{eqnarray}
In the field-free case, $u_{\hat{\bm{ q}}}$ is isotropic (in the bcc
crystal), but in the magnetic field it becomes anisotropic.

\begin{figure}[!t]
\begin{center}
\leavevmode
\includegraphics[height=90mm,bb=71 455 538 741,clip]{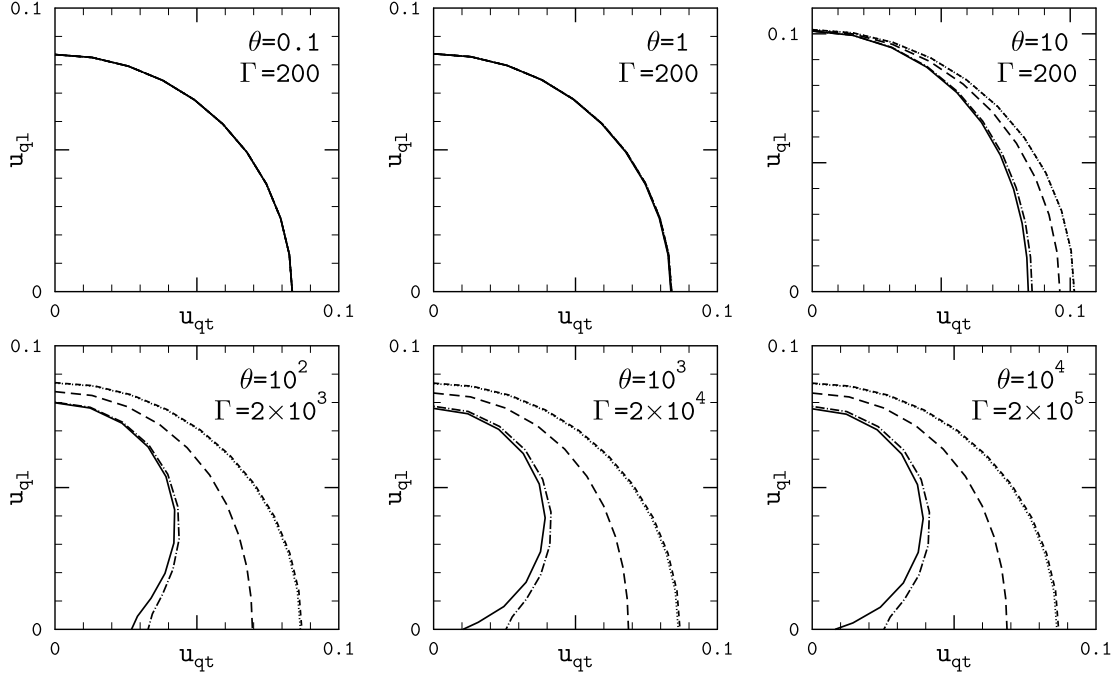}
\end{center}
\vspace{-0.3cm} \caption[ ]{The rms ion displacement from the
equilibrium position (the origin). $Y$-axis ($u_{\rm ql}$) is
parallel, $X$-axis ($u_{\rm qt}$) is perpendicular to ${\bm B}$.
$u_{\rm ql}$ and $u_{\rm qt}$ are in units of the distance to the
nearest neighbor. The solid, dash-dotted, long-dashed, dotted, and
short-dashed curves correspond to $b = 100, 10, 1, 0.1$, and 0.01,
respectively. The distance between the origin and a point on a
chosen curve is equal to the rms ion displacement $u_{\hat{\bm{
q}}}$ in that direction. }
\label{dw}
\end{figure}
%

Amplitudes of ion oscillations are depicted in Fig.\ \ref{dw}. The
ion equilibrium position is at the origin. The $Y$-axis is parallel
to $\bm{B}$ and to the direction towards one of the nearest
neighbors. There is symmetry in the plane azimuthal with respect to
$\bm{ B}$ (the plane perpendicular to the $Y$-axis). Hence, the
$X$-axis indicates an arbitrary direction orthogonal to $\bm{ B}$.
The distance between the origin and a point on a chosen curve is
equal to the rms displacement $u_{\hat{\bm{ q}}}$ in the given
direction. The distance is expressed in units of the spacing between the
nearest neighbors, $a_{n} =  a_{l} \sqrt{3} /2 = (3 \pi^2)^{1/6} a$
(for bcc lattice). Solid, dash-dotted, long-dashed, dotted, and
short-dashed curves correspond to $b = 100, 10, 1, 0.1$, and 0.01,
respectively. The 6 panels of Fig.\ \ref{dw} are for the 6 values of
the quantum parameter $\theta$, from $0.1$ to $10^4$. The ion
displacement at a different $\Gamma$ (for given $\theta$ and $b$)
can be found by straightforward scaling per Eq.\ (\ref{r2T}).

The quantity $u_{\hat{\bm{ q}}}$ is related to the Debye-Waller
factor $W(\hat{\bm{ q}})$: $\langle \exp(i \hat{\bm{ q}} \cdot
\hat{\bm{ u}}) \rangle_T = \exp(- W)$. The thermal averaging yields
\begin{equation}
      \langle \exp{\left(i \hat{\bm{ q}} \cdot \hat{\bm{ u}} 
         \right)}\rangle_T =
          \exp{\left[-\frac{\hbar^2}{MN}
       \sum_{\bm{k}s} \hat{q}^\lambda \hat{q}^\mu
         \alpha_{\bm{k}s}^{\lambda \ast} \alpha_{\bm{k}s}^\mu
    \left(\bar{n}_{\bm{k}s}+ \frac{1}{2} \right) \right]}~.
\nonumber
\end{equation}
Thus, $u^2_{\hat{\bm{ q}}} = 2W(\hat{\bm{ q}})$.

As shown in Fig.\ \ref{dw}, the ion displacements at $\theta \lesssim 1$
are insensitive to the magnetic field (all 5 lines merge) and
isotropic. At $\theta \sim 10$ the displacement along the field
remains essentially the same as at $B=0$, whereas the
transverse displacements shrink. The net effect will be an increase of the
crystal stability and melting temperature. The anisotropy of
displacements in a quantum, strongly magnetized crystal, at $\theta
\gtrsim 100$, $b \gg 1$, becomes much larger. The ion displacements
are strongly suppressed in the transverse direction and are mildly
reduced along the field (see also \cite{NF83}). 

These effects are
important for accurate computations of nuclear reaction rates in a
strongly magnetized crust of a neutron star (Sec.\ \ref{introduct}).
The strongest effect is expected to occur at sufficiently high
densities and not too high temperatures in the pycnonuclear burning
regime. In that regime (e.g., \cite{SVH69,YGABW06,CD09}) the main
contribution to the reaction rate comes from the closest neighbors
in a crystalline lattice due to their zero-point vibrations.
In the field-free case, the rate depends exponentially on the squared
ratio of the equilibrium distance between the closest neighbors and
the average amplitude of their displacements. The ratio is typically
large, and the reaction rates are exponentially small (but increase with
the growth of stellar matter density). Although the reaction rates
in magnetized crystals have not been studied yet, one can expect
a similar result, involving now the average displacement (\ref{u2q})
in the direction of closest neighbors. If so, the rates
would become extremely sensitive to the orientation of the magnetic
field. If the field is not directed towards one of the nearest
neighbors, the reactions will be significantly slowed down by greater
tunnelling lengths. Even if the magnetic field is directed towards
one of the nearest neighbors then, in addition to the reduction of
the longitudinal displacement, there will be a geometrical effect.
In a bcc lattice every ion has 8 nearest neighbors, of which only 2
will be located along the magnetic field line. The reactions with
the other 6 neighbors will be quenched.

\section{Phonon spectrum moments in the magnetic field}
\label{moments}
According to the well-known Bohr-van Leeuwen theorem,
the classical partition function is
not affected by the magnetic field. For instance,
the classical asymptote of the entropy $S$ has the form:
\begin{equation}
    \frac{S}{3nV} = 1 -  \ln{\left( \frac{\hbar \omega_{p}}{T} \right)}
     -  \left\langle \ln{\left(
   \frac{\Omega}{\omega_{p}} \right)} \right\rangle_{\rm ph}~.
\end{equation}
Therefore, the quantity $\langle
\ln{(\Omega/\omega_{p})}\rangle_{\rm ph}$, where average over all
phonon modes in the first Brillouin zone is implied, {\it should not
depend} on the magnetic field. This statement is easy to prove
directly. By inspecting Eq.~(\ref{secular-2}) one concludes that 
its $\Omega^0$-term cannot contain $B$ at any $\bm{k}$. On the other hand,
this term is equal to $\Omega_1^2 \Omega_2^2 \Omega_3^2$, 
which proves the point.

\begin{figure}[ht]
\begin{center}
\leavevmode
\includegraphics[height=73mm,bb=71 532 538 740,clip]{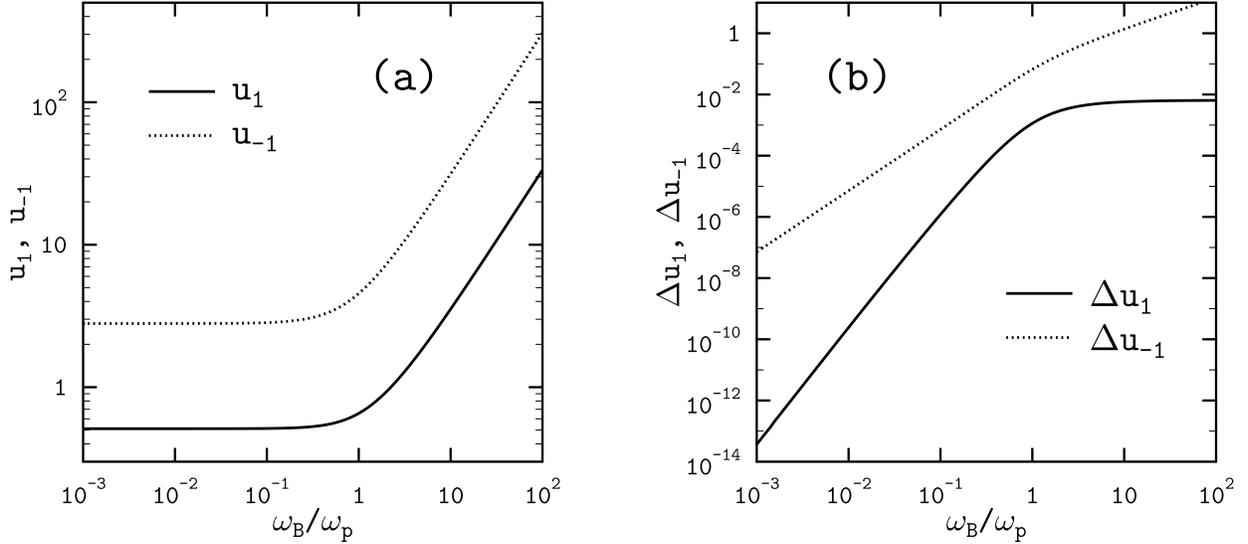}
\end{center}
\vspace{-0.2cm} \caption[ ]{(a): The dependence of phonon 
spectrum moments $u_1$ and $u_{-1}$
on the strength of the magnetic field directed towards
one of the nearest neighbors. (b): The dependence 
of phonon spectrum moment differences 
$\Delta u_1$ and $\Delta u_{-1}$ on the magnetic field 
(see explanation in the text).}
\label{u1um1}
\end{figure}

Phonon frequency moments $u_m \equiv \langle (\Omega /\omega_{p})^m
\rangle_{\rm ph} = u_m(b,\bm{n})$ depend on both 
the field strength and direction $\bm{n}$.
The moment $u_{-2}$ diverges, while the moment $u_2 = 1 + b^2$.
Consider the moments $u_1$ and $u_{-1}$. Their dependences on the
field strength are shown in Fig.\ \ref{u1um1}(a) for the field
direction corresponding to the minimum of the zero-point energy. At
strong magnetic fields both moments behave in the same way,
proportional to $b$, although under the effect of different modes.
The main contribution to $u_1$ comes from the mode $\Omega_3 \approx
\omega_B$, whereas the major contribution to $u_{-1}$ is due to the mode
$\Omega_1 \propto 1/B$ (near the center of the Brillouin zone).

%
%

The dependence of the phonon frequency moments
on the magnetic field direction
turns out to be rather weak.
In Fig.\ \ref{u1um1}(b) the differences
$\Delta u_1 = u_1(b,\bm{n}_2) - u_1(b,\bm{ n}_1)$
and $\Delta u_{-1} = u_{-1}(b,\bm{ n}_2) - u_{-1}(b,\bm{ n}_1)$ are shown,
where $\bm{n}_{1}$ is the magnetic field direction
towards one of the nearest neighbors, while
$\bm{n}_{2}$ is the direction towards one of the next order nearest
neighbors.

The behavior of $\Delta u_1$ is of special importance because it
determines zero-point energy gain $\Delta E_{\rm zp} = \frac32 N \hbar
\omega_{p} \Delta u_1$ resulting from the crystal rotation with
respect to the magnetic field. In weak fields $10^{-3} \lesssim b <
1$, the difference $\Delta u_1$ scales approximately as $b^4$. In
strong fields, $\Delta u_1$ saturates at a value $\sim 10^{-2}$,
depending on the field direction. The difference $\Delta u_{-1}$
scales as $b^2$ for low magnetic fields, and as $b$ for strong magnetic
fields.

\section{Conclusion}
\label{concl}
The Coulomb crystal of ions with incompressible charge 
compensating background
of electrons in constant uniform 
magnetic field has been studied. The phonon mode spectrum
of the crystal with bcc lattice has been calculated 
for a wide range of magnetic field strengths and orientations
(Fig.\ \ref{spectinb}). 
The phonon spectrum has been used in 3D numerical integrations
over the first Brillouin zone for a detailed quantitative analysis of
the phonon contribution to the crystal thermodynamic functions, 
Debye-Waller factor of ions, and the rms ion displacements from
the lattice nodes for a broad range of densities, temperatures, chemical
compositions, and magnetic fields. 

The characteristic parameter that determines the strength of the
magnetic field effects in the crystal is the ratio of the 
ion cyclotron frequency to
the ion plasma frequency, $b = \omega_B/\omega_{p}$. Even a very
strong field ($b \gg 1$) does not alter the partition function
(hence, thermodynamic functions, melting etc.) of a classical
crystal, i.e.\ a crystal at a temperature significantly exceeding
the energy of any phonon mode 
($T \gg \hbar \omega_B$ for strongly magnetized crystal).

Strong magnetic field dramatically changes various properties of
quantum crystals, especially at $\theta = \hbar \omega_{p}/T \gg 1$.
Low-temperature phonon specific heat, entropy, thermal energy increase 
by orders of magnitude (e.g., Figs.\ \ref{b-c}, \ref{b-se}). The 
thermodynamic
functions exhibit peculiar staircase structures, brought about by the vastly
different energy scales of the crystal phonon mode. 
Ion displacements from the equilibrium
positions become strongly anisotropic (Fig.\ \ref{dw}), and so does
the Debye-Waller factor. 

These results can be applied directly to real physical systems found
in the crust of magnetars (neutron stars with superstrong magnetic field).
In particular, the heat capacity of the magnetized Coulomb crystal 
composed of
fully ionized $^{56}$Fe has been analyzed (Sec.\ \ref{crust_ex}).
It has been shown that in the magnetic field (as in the field-free case)
the phonon heat capacity dominates the electron one
in a wide range of densities and temperatures 
(Fig.\ \ref{iron_spec_heat}). In accordance with Sec.\ \ref{thermodyn},
magnetic field is responsible for a significant boost of phonon heat 
capacity,
which may be important for simulations of magnetar cooling.
Since ion displacements are suppressed, one
can expect increased crystal stability in the magnetar crust
associated with an increase of the melting temperature in the quantum regime. 
Finally, one expects a reduction of the pycnonuclear reaction 
rates due to increased
tunnelling lengths. The rates are likely to become very sensitive to
the orientation of the magnetic field (Sec. \ref{rms}).

The dependence of thermodynamic functions on the magnetic field
orientation within the crystal turns out to be weak, at least for
the bcc lattice, but finite. The energy gain, achieved by
reorienting the crystal with respect to the magnetic field, can be
estimated using Fig.\ \ref{u1um1}.

The techniques used in the present paper can be easily 
reformulated for fcc lattice,
where results are expected to be numerically close to those for bcc lattice.

\acknowledgments{The author is deeply grateful to D.G.\ Yakovlev for
numerous discussions and valuable comments to the manuscript. 
This work was partly
supported by the Russian Foundation for Basic Research (grant
08-02-00837) and by the State Program ``Leading Scientific Schools
of Russian Federation'' (grant NSh 2600.2008.2).}

\end{document}